\documentclass[twocolumn]{aastex61_arxiv}

\usepackage{multirow}
\usepackage{amsmath}

\newcommand{\msun}{\ensuremath{M_\sun}}

\submitjournal{ApJL}

\shorttitle{Chromium in Si-C Shell Mergers}
\shortauthors{C\^ot\'e et al.}

\begin{document}

\title{Chromium Nucleosynthesis and Silicon-Carbon Shell Mergers in Massive Stars}

\correspondingauthor{Benoit C\^ot\'e}
\email{benoit.cote@csfk.mta.hu}

\author[0000-0002-9986-8816]{Benoit C\^ot\'e}
\affiliation{Konkoly Observatory, Research Centre for Astronomy and Earth Sciences, Hungarian
Academy of Sciences, Konkoly Thege Miklos ut 15-17, H-1121 Budapest, Hungary}
\affiliation{National Superconducting Cyclotron Laboratory, Michigan State University, East
Lansing, MI 48824, USA}
\affiliation{Joint Institute for Nuclear Astrophysics - Center for the Evolution of the
Elements, USA}
\affiliation{NuGrid Collaboration, \url{https://nugrid.github.io}}

\author[0000-0003-3970-1843]{Samuel Jones}
\affil{X Computational Physics (XCP) Division, Los Alamos National Laboratory, Los Alamos, New Mexico 87545, USA}
\affiliation{NuGrid Collaboration, \url{https://nugrid.github.io}}

\author[0000-0001-8087-9278]{Falk Herwig}
\affiliation{Department of Physics and Astronomy, University of Victoria, Victoria, Canada}
\affiliation{Joint Institute for Nuclear Astrophysics - Center for the Evolution of the
Elements, USA}
\affiliation{NuGrid Collaboration, \url{https://nugrid.github.io}}

\author[0000-0002-9048-6010]{Marco Pignatari}
\affiliation{E.A. Milne Centre for Astrophysics, Department of Physics \& Mathematics, University of Hull, HU6 7RX, United Kingdom.}
\affiliation{Konkoly Observatory, Research Centre for Astronomy and Earth Sciences, Hungarian Academy of Sciences, Konkoly Thege Miklos ut 15-17, H-1121 Budapest, Hungary}
\affiliation{Joint Institute for Nuclear Astrophysics - Center for the Evolution of the
Elements, USA}
\affiliation{NuGrid Collaboration, \url{https://nugrid.github.io}}

\begin{abstract}

We analyze the production of the element Cr in galactic chemical evolution (GCE) models using the NuGrid nucleosynthesis yields set. We show that the unusually large [Cr/Fe] abundance at [Fe/H]\,$\approx 0$ reported by previous studies using those yields and predicted by our Milky Way model originates from the merging of convective Si-burning and C-burning shells in a 20\,\msun~model at metallicity $Z=0.01$, about an hour before the star explodes. This merger mixes the incomplete burning material in the Si shell, including $^{51}$V and $^{52}$Cr, out to the edge of the carbon/oxygen (CO) core. The adopted supernova model ejects the outer 2\,\msun~of the CO core, which includes a significant fraction of the Cr-rich material. When including this 20\,$M_\odot$ model at $Z=0.01$ in the yields interpolation scheme of our GCE model for stars in between 15 and 25\,\msun, we overestimate [Cr/Fe] by an order of magnitude at [Fe/H]\,$\approx$\,0 relative to observations in the Galactic disk. This raises a number of questions regarding the occurrence of Si-C shell mergers in nature, the accuracy of different simulation approaches, and the impact of such mergers on the pre-supernova structure and explosion dynamics. According to the conditions in this 1D stellar model, the substantial penetration of C-shell material into the Si-shell could launch a convective-reactive global oscillation, if a merger does take place. In any case, GCE provides stringent constraints on the outcome of this stellar evolution phase.

\end{abstract}


\section{Introduction}
\label{sec:intro}

Galactic chemical evolution (GCE) models and simulations are powerful tools to bridge nuclear astrophysics with astronomical observations (e.g., \citealt{1980FCPh....5..287T,2003PASA...20..401G,2014SAAS...37..145M,2013ARA&A..51..457N,2018MNRAS.476.3432P}).  In spite of the complexity associated with simulating the formation and evolution of galaxies (e.g., \citealt{2012ApJ...745...50W,2015MNRAS.446..521S,2015ARA&A..53...51S,2018MNRAS.480..800H,2018MNRAS.473.4077P,2018A&A...616A..96R}), the fundamental input ingredients of all GCE studies are still the stellar yields (e.g., \citealt{2010A&A...522A..32R,2015MNRAS.451.3693M,2018ApJ...861...40P}). In the past years, we have developed an open-source galactic chemical evolution (GCE) pipeline in order to bring nuclear astrophysics efforts to the forefront of GCE studies.

There are several sources of uncertainties in generating grids of stellar models
for GCE applications, including, for example, uncertainties in nuclear reaction
rates (e.g.,
\citealt{2004ApJ...615..934L,2009ApJ...702.1068T,2014ApJ...795..141T,2017RvMP...89c5007D,2017MNRAS.469.1752N,2018JPhG...45e5203D,2018ApJS..234...19F}),
stellar evolution and internal mixing (e.g.,
\citealt{2007ApJ...667..448M,Sukhbold2014a,jones2015a,Davis2019a}), and
supernova explosion modeling (e.g.,
\citealt{Sukhbold2016a,2018ApJ...856...63F,2019arXiv190201340C,2019ApJ...870....1E,2019arXiv190204270M}).  Turning this
argument around, GCE studies are ideal framework to explore the impact of
stellar processes in a broader astronomical context (\citealt{2017ApJ...835..128C}). In this study, we focus on
the impact of shell mergers occurring in NuGrid massive star models
(\citealt{2018MNRAS.480..538R}) during the pre-supernova evolution phase (see
also \citealt{Rauscher2002a}, \citealt{2018MNRAS.481.2918M}, and \citealt{Yadav2019a}). 

\cite{Ritter2018a} have shown that oxygen-carbon (O-C) shell mergers could
potentially be a relevant site for the production of odd-Z elements and
p-process isotopes at galactic scale. \citet{Andrassy2018a}, motivated by this,
have studied the 3D hydrodynamical properties of O-C shell mergers. Here we
discuss the impact of silicon-carbon (Si-C) shell mergers on the evolution of
chromium (Cr) in the Milky Way. Since the publication of the second set of
NuGrid yields (\citealt{2018MNRAS.480..538R}), an overproduction of Cr at
galactic scale has been reported by \cite{2018ApJS..236....2H} and
\cite{2018ApJ...861...40P} when using these yields in GCE codes. We have
isolated the source of this overproduction. In the 20\,$M_\odot$ model at
$Z=0.01$, a Si-C shell merger mixes large amounts of Cr, synthesized during
Si-shell burning, above the assumed mass cut\footnote{Anything below the mass
cut is locked inside the compact remnant and does not contribute to the
ejected yields.}.

In Section~\ref{sec:gce} we use our chemical evolution tools to highlight the Cr
overproduction that points to the specific stellar model responsible for this
overproduction. In the other subsections of Section~\ref{sec:results}, we present the relevant parts of the stellar model, show the implication of the Si-C shell merger on the pre-supernova structure, and discuss the convective and burning time scales during the merger event.
In Section~\ref{sec:conclusions}, we conclude and raise open questions regarding the occurrence of Si-C shell mergers in nature and in multi-dimensional hydrodynamic simulations.

\section{Results}
\label{sec:results}

\subsection{Galactic Chemical Evolution Model}
\label{sec:gce}
We use the GCE code \texttt{OMEGA+} (\citealt{2018ApJ...859...67C}) to bring
NuGrid yields (\citealt{2018MNRAS.480..538R}) into a galactic context. This code
is part of the open-source JINAPyCEE python
package\footnote{\url{https://github.com/becot85/JINAPyCEE}} and represents a
one-zone GCE model surrounded by a large circumgalactic gas reservoir. 
The input parameters adopted in this study for regulating the star formation efficiency and the galactic inflow and outflow rates are available
online\footnote{\url{https://github.com/becot85/JINAPyCEE/blob/master/DOC/OMEGA\%2B_Milky_Way_model.ipynb}}.
We use the initial mass function of \cite{2001MNRAS.322..231K}. For Type~Ia supernovae we use the yields of \cite{1999ApJS..125..439I} and assume a total number of $10^{-3}$ Type~Ia supernova per unit of stellar mass formed.
As shown below, the overproduction of Cr is so strong that the choice of
GCE parameters and Type~Ia supernova yields is of little
importance.

\begin{figure}
\center
\includegraphics[width=3.35in]{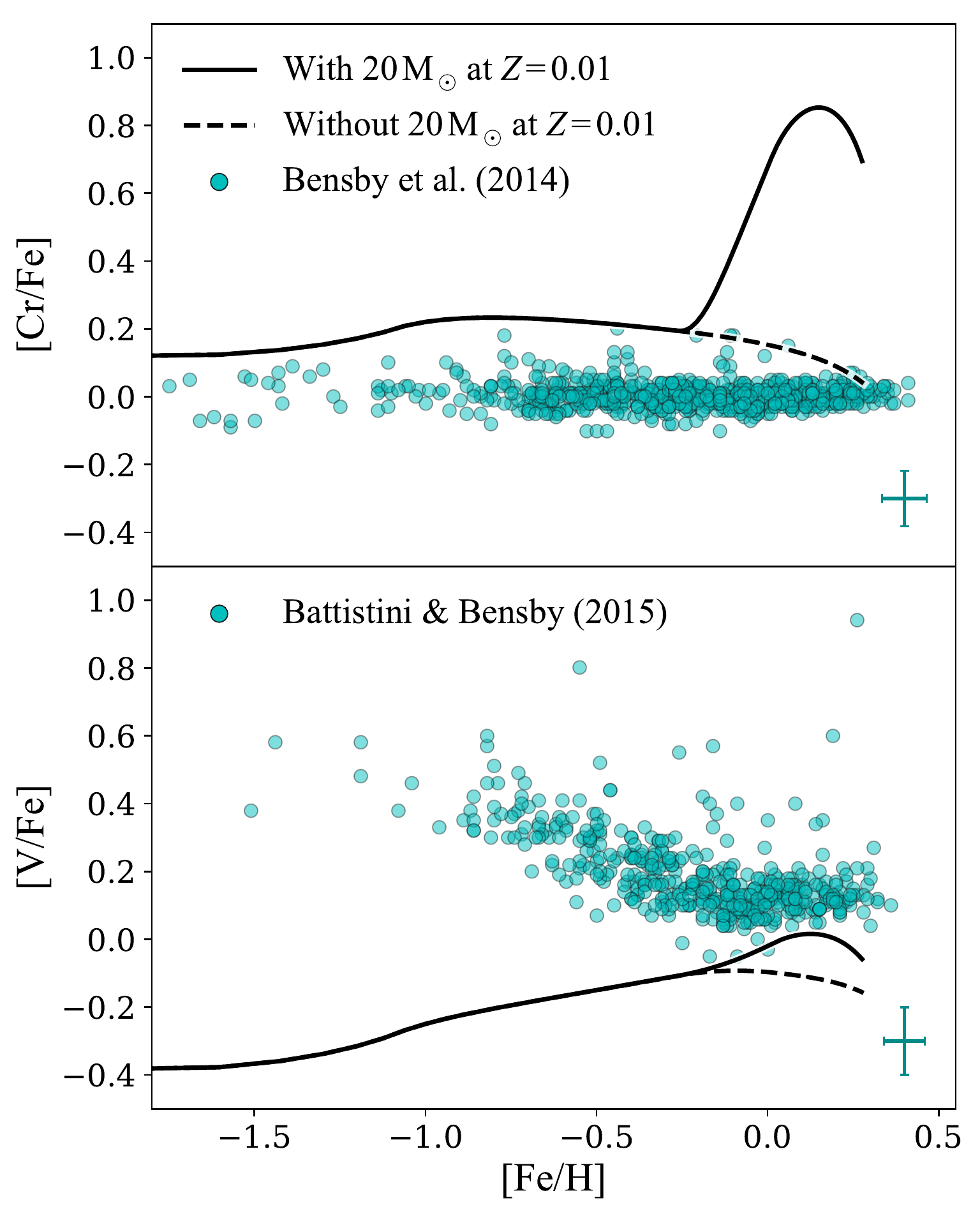}
\caption{Predicted evolution of [Cr/Fe] (top panel) and [V/Fe] (bottom panel) as a function of [Fe/H] for the Galactic disk using NuGrid yields (\citealt{2018MNRAS.480..538R}) and the GCE code \texttt{OMEGA+} (\citealt{2018ApJ...859...67C}). The solid and dashed lines show the predictions when including or excluding the 20\,M$_\odot$ model at $Z=0.01$, respectively. The cyan dots are stellar disk data from \cite{2014A&A...562A..71B} and \cite{2015A&A...577A...9B}.
\label{fig:gce}}
\end{figure}

The upper panel of Figure~\ref{fig:gce} shows the predicted evolution of Cr
abundances as a function of [Fe/H].
Near [Fe/H]\,$=$\,$0$, our predictions (solid line) have a bump that
overestimates disk data by almost an order of magnitude.  The 20\,$M_\odot$ stellar model at $Z=0.01$ is at the origin of the
Cr bump (see also Figure~7 in \citealt{2018ApJS..236....2H}). When
removing this stellar model from the yields set, the bump disappears entirely (dashed line in Figure~\ref{fig:gce}). The bottom panel shows a production of V accompanying the production of Cr. This is not surprising, since V and Cr are made efficiently at similar stellar conditions \citep[e.g.,][]{woosley:95}.
In agreement with the simulations reported here, [V/Fe] is typically 
underestimated in GCE models at all metallicities compared to observations (e.g., \citealt{2018MNRAS.476.3432P} and references therein).

\begin{figure*}
	\includegraphics[width=\textwidth]{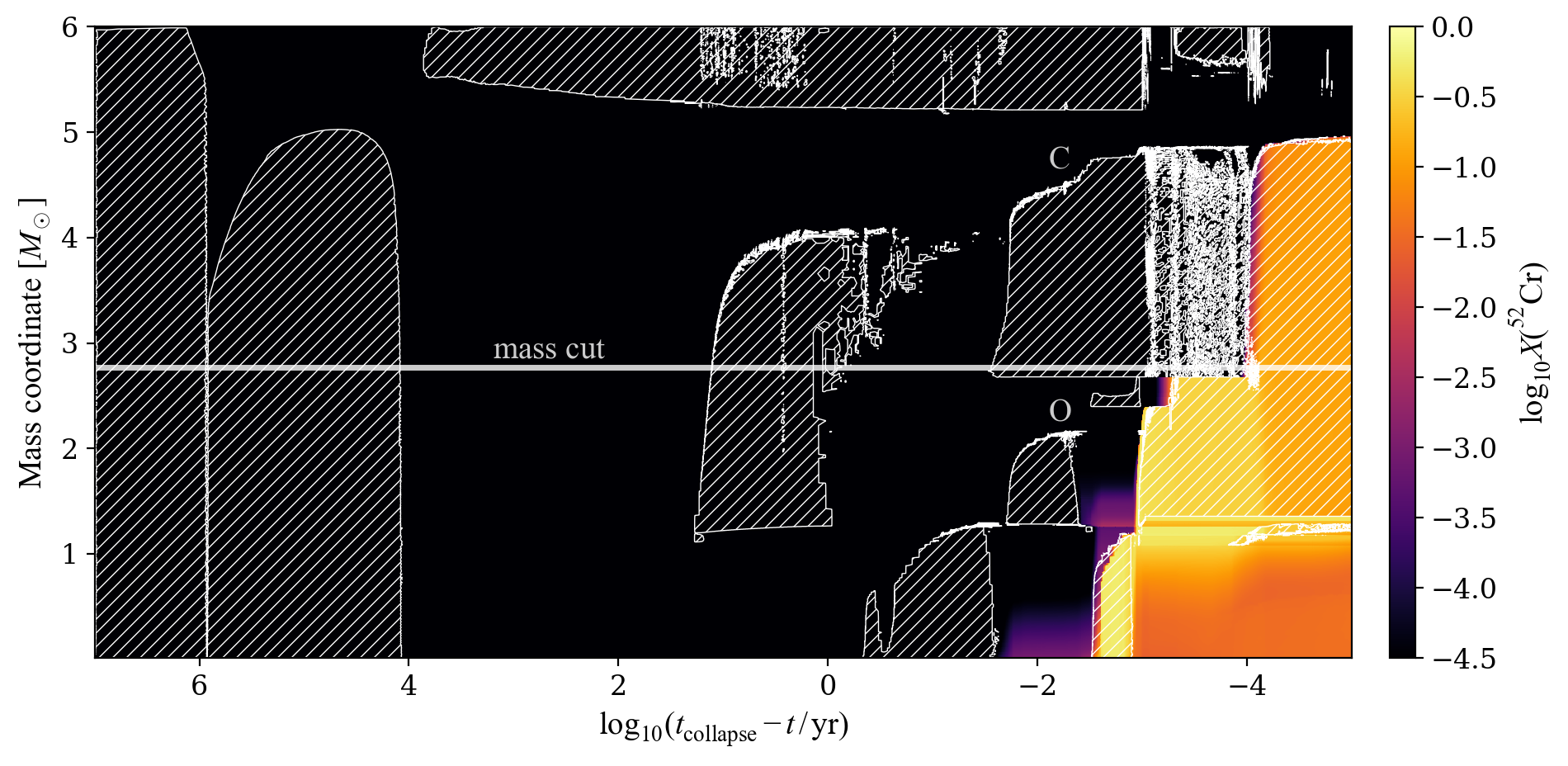} \\
	\includegraphics[width=\textwidth]{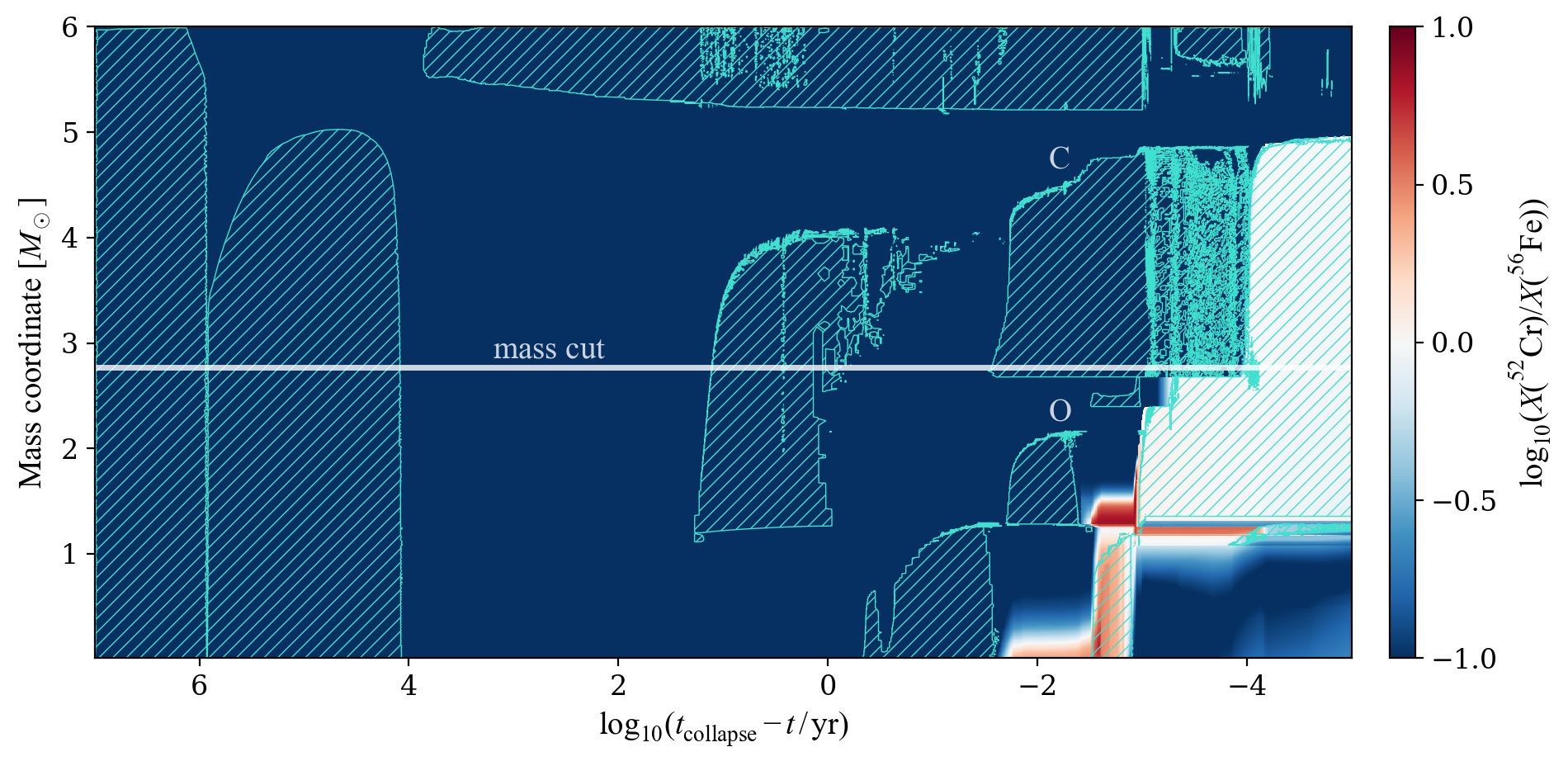}
	\caption{Kippenhahn (convective structure evolution) diagrams of the
		inner 6\,\msun~of the 20\,\msun~model at $Z=0.01$.
		The x-axis has an
		inverse logarithmic time scale showing the time remaining before collapse. Hatched
		contours show convectively unstable regions and color indicates
		the mass fraction of $^{52}$Cr in the top panel and the
		$^{52}$Cr/$^{56}$Fe ratio in the bottom panel. The ratio is
		approximately unity in the merged C/O/Si shell at the presupernova
		stage, which is the signature of incomplete Si burning that has
		been mixed throughout the merged shells. This is the predominant
		signature that appears in the ejected yields from the stellar
		model.
	}
	\label{fig:kips}
\end{figure*}

\subsection{Final Evolutionary Stages of the Stellar Model}
Here we describe the evolution of the 20\,\msun~model at $Z=0.01$. We show that the source of Cr comes from the merging of the C- and Si-burning convective shells during the last hour of
evolution before the model collapses.

The Kippenhahn diagrams in Figure~\ref{fig:kips} show the evolution of convection
zones (hatched regions) in the inner 6\,\msun~of the model. At 
$t^* \equiv (t_\mathrm{collapse} - t) / \mathrm{yr} \approx 10^{-3}$, 
the convective Si-burning shell engulfs the overlying radiative layer that separated it from the base of the convective C burning shell above. From that time until
$t^* \approx 10^{-4},$
the stratified Si- and C-burning convective shells share a
convective boundary, which prevents the transport of material from the C
shell into the Si shell, and vice versa. Finally, at
$t^* \approx 10^{-4},$
the convective boundary is eroded and the Si burning shell extends from the edge
of the Fe core almost to the edge of the CO core. The top panel of
Figure~\ref{fig:kips} shows that $^{52}$Cr is mixed out to the edge of the CO
core and its mass fraction is slightly reduced owing to dilution by the C shell
material.

As shown by our GCE model (Figure~\ref{fig:gce}), a boost of [Cr/Fe] is generated above [Fe/H]\,$\approx-0.2$ when the yields of this 20\,\msun~model are included. One might expect that if
the Si-burning shell was mixed out and ejected, then [Cr/Fe] should stay
relatively flat, since both Cr and Fe are predominant products of Si burning.
However, because the Si shell merges while it is still convective, and hence
still burning, the ejected chemical signature is typical of incomplete Si burning.

\begin{figure}
\center
\includegraphics[width=3.35in]{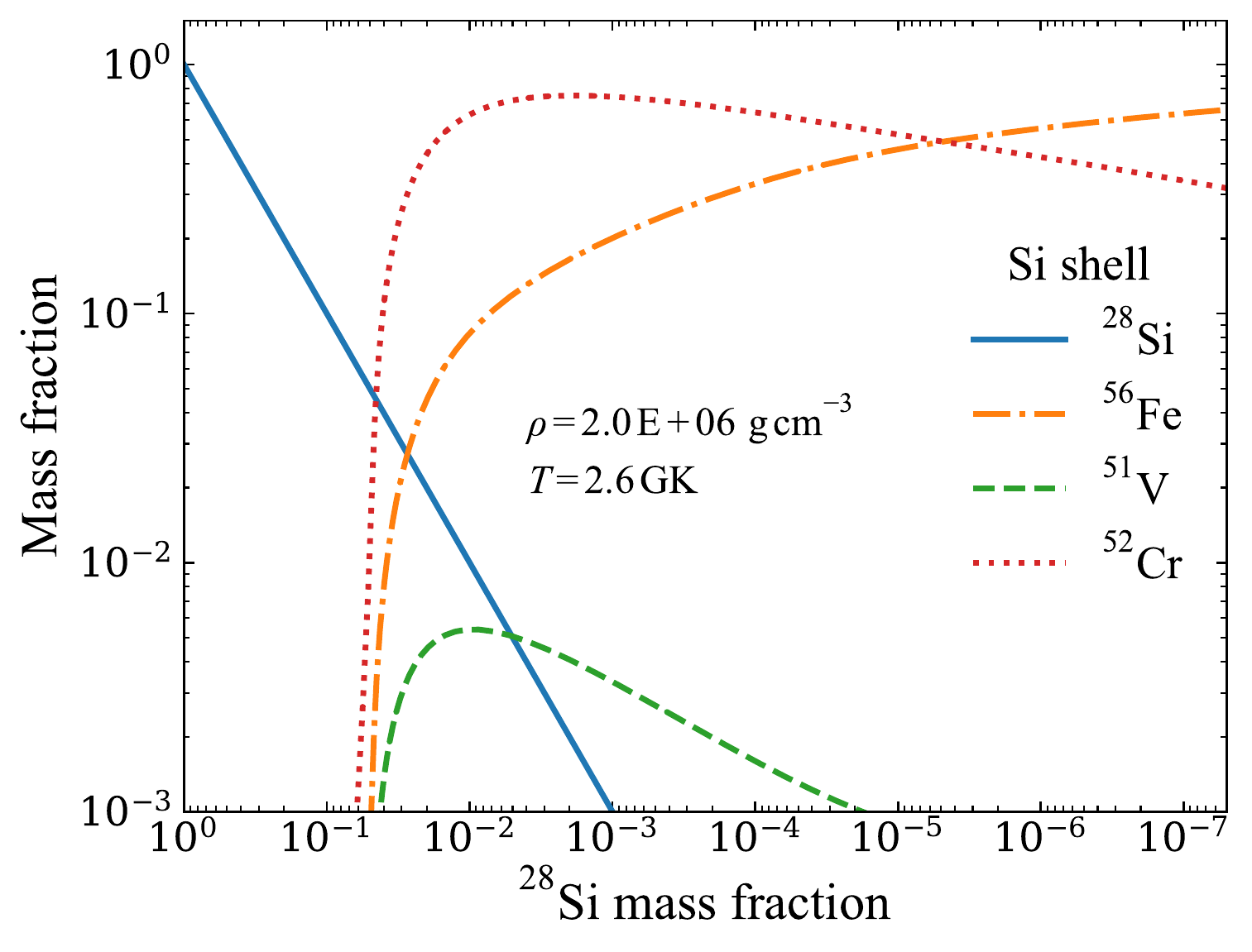}
\caption{Evolution of key isotopes during a one-zone burn network integration at
	conditions characteristic of Si shell burning in a 20~\msun~star.
\label{fig:burn}}
\end{figure}

Si burning 
produces the neutron-magic isotope $^{52}$Cr relatively
quickly, and thereafter produces $^{56}$Fe via a sequence of capture and
photodisintegration processes. This is illustrated in Figure~\ref{fig:burn},
which shows the results of a one-zone nuclear reaction network calculation
starting from pure $^{28}$Si and evolved at density $\rho = 2\times10^6$~g~cm$^{-3}$
and temperature $T = 2.6$~GK, which are approximately the conditions during shell Si
burning in the 20\,\msun~stellar model about an hour before collapse. We note
that the time scales of the burning will be different in the star owing to the
evolution of the core acting to increase the density and temperature of the
shell. However, we have confirmed that the behaviour is similar for a range of temperatures and densities.  Only in the most extreme conditions ($10^{10}$~g\,cm$^{-3}$, 5~GK) is Cr produced at the same time as Fe.

The core-collapse supernova (CCSN)
explosion for this star was modelled by \citet{2018MNRAS.480..538R} in the same
way as described in \citet{pignatari2016a}, with a mass cut at 2.77\,\msun~based
on remnant mass prescription derived in \citet{fryer2012a}. In this particular
model with these assumptions, a significant fraction of the incomplete Si-shell
burning material, including $^{52}$Cr and $^{51}$V, is ejected during the
explosion (the convective Si-C shell extends up to $\sim$\,2\,\msun~above the
assumed mass cut). The yields from the stellar model are therefore heavily influenced by the
occurrence of a Si-C shell merger and by the methods employed to model the
supernova. 

\subsection{Supernova Implications}

The 2.77\,\msun~compact remnant is a black hole created by fallback
\citep{fryer2012a}.
In agreement with \cite{Sukhbold2016a} and \cite{2019ApJ...870....1E}, stars around 20\,\msun~are generally found to produce more failed explosions and black holes in 1D simulations than for lower initial masses.
However, CCSN theory has far from settled on the precise
relationship between progenitor structure and explosion properties
\citep{Sukhbold2014a,Mueller2016a,2018ApJ...856...63F}. Even if we did know the
remnant mass for a typical model of this initial mass and metallicity, 
the structure of the core in our model is
affected by the shell merger in ways that are likely important in determining
the dynamics of the collapse and subsequent explosion, or lack thereof, and
compact remnant mass (e.g., \citealt{Davis2019a,Yadav2019a}).

\begin{figure*}
	\centering
	\includegraphics[width=\textwidth]{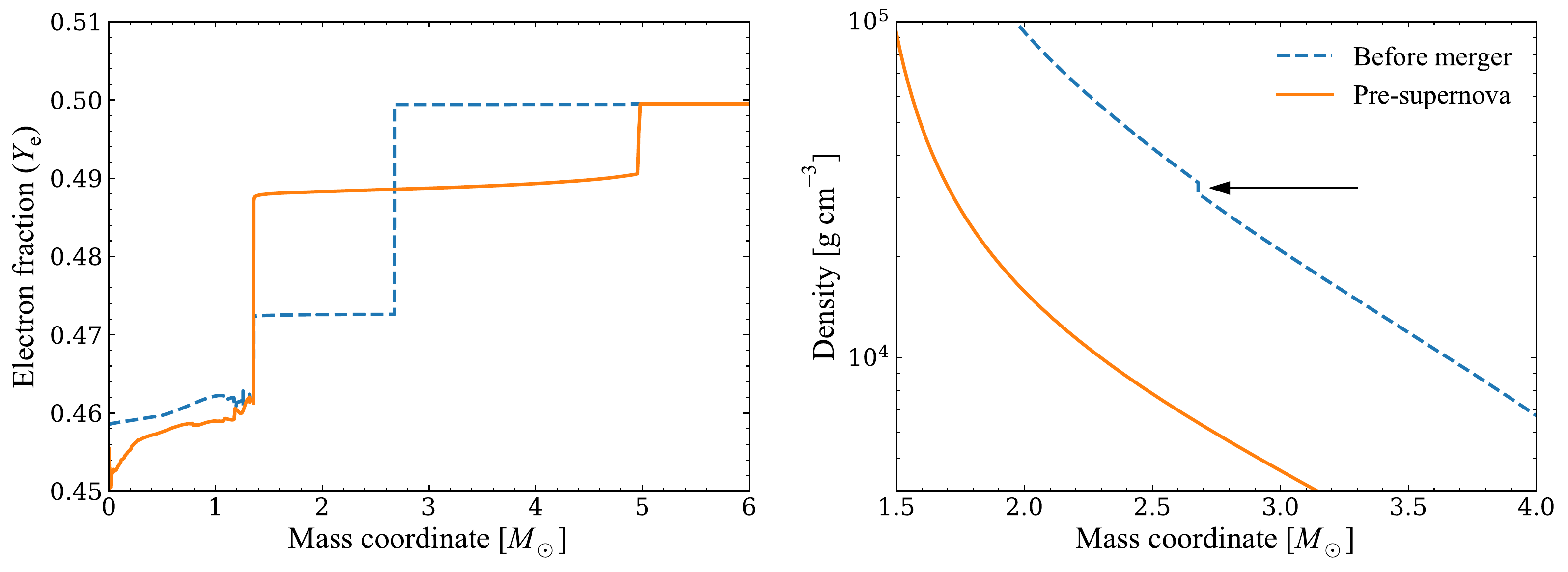}
	\caption{Electron fraction profile (left panel) and density profile
		(right panel) in the core of the stellar model
	before the shell merger and at the pre-supernova stage.}
	\label{fig:ye}
\end{figure*}

Two of the most important properties of the progenitor models for determining
the outcome of the CCSN are the stratification of density and electron fraction
($Y_\mathrm{e}$). The left panel of Figure~\ref{fig:ye} shows how the
$Y_\mathrm{e}$ profile is altered by the merger. The right panel shows that the
shell merger erases the density jump that existed at the interface of the two
shells (see arrow). Such a jump may facilitate the revival of the stalled CCSN
shock wave owing to the rapid drop in accretion rate as the shell arrives at the
shock radius \citep{Ott2018a}. Although this particular density jump appears
small, it could still be enough to alleviate the ram pressure at a critical time
and allow for a successful explosion.  Conversely, it may be that a more
realistic simulation of this progenitor model results in direct black hole
formation or formation of a much larger black hole than 2.77~\msun~by fallback,
in which case we would perhaps expect none of the CO core (and hence, none of
the Cr) to be ejected. Further implication for the SN explosion may derive from asymmetries that could be seeded right before the explosions in a shell merger, depending on the time scales for convection and burning during the merger.
\\
\\
\subsection{Convective Time Scale}
The $Y_\mathrm{e}$ profile shown in Figure~\ref{fig:ye} raises another
interesting point. The profile is not flat in the newly combined convection zone
between 1.4 and 5~\msun~(see solid orange line),
as revealed by the time-dependent mixing, implemented in the diffusion approximation, when nuclear and mixing time-scales are similar.
The pre-supernova profile represents a state of incomplete mixing
of the material in the two shells.
The convective time
scale $\tau_\mathrm{conv}$, assuming it is approximately the time taken for a
fluid element to complete one cycle of advection around a convective cell whose
diameter is the shell's thickness, is given by
\begin{equation}
	\tau_\mathrm{conv} \approx \dfrac{2\pi r_\mathrm{cell}}{\langle
	v_\mathrm{conv}\rangle} \approx
	\dfrac{(7.4\times10^9)
	\pi\;\mathrm{cm}}{6.3\times10^6\;\mathrm{cm~s}^{-1}} = 3690 \;
	\mathrm{s}.
\end{equation}
The shell merger takes place $10^{-4}$~yr or 3154 seconds before collapse, which
is a similar time scale.

This raises the question of how efficient will be the mixing of $^{52}$Cr and
other Si-burning products into the outer core if there is only one turnover time
to do it. Certainly the use of mixing length theory (MLT) for convection becomes
inappropriate under these conditions because MLT only predicts convection properties in terms of averages over many convective turn-over time-scales.

\subsection{Burning Time Scale During Shell Merger}
\label{sec_da}
The constitution of the C shell should burn rapidly when exposed to the
temperatures in the Si-burning shell. This energetic feedback will likely modify
the flow dynamics and it should be considered when modeling the pre-supernova
evolution of such a star \citep[e.g.][]{Herwig2014a,Andrassy2018a,Yadav2019a}.

\begin{figure}
	\centering
	\includegraphics[width=\linewidth]{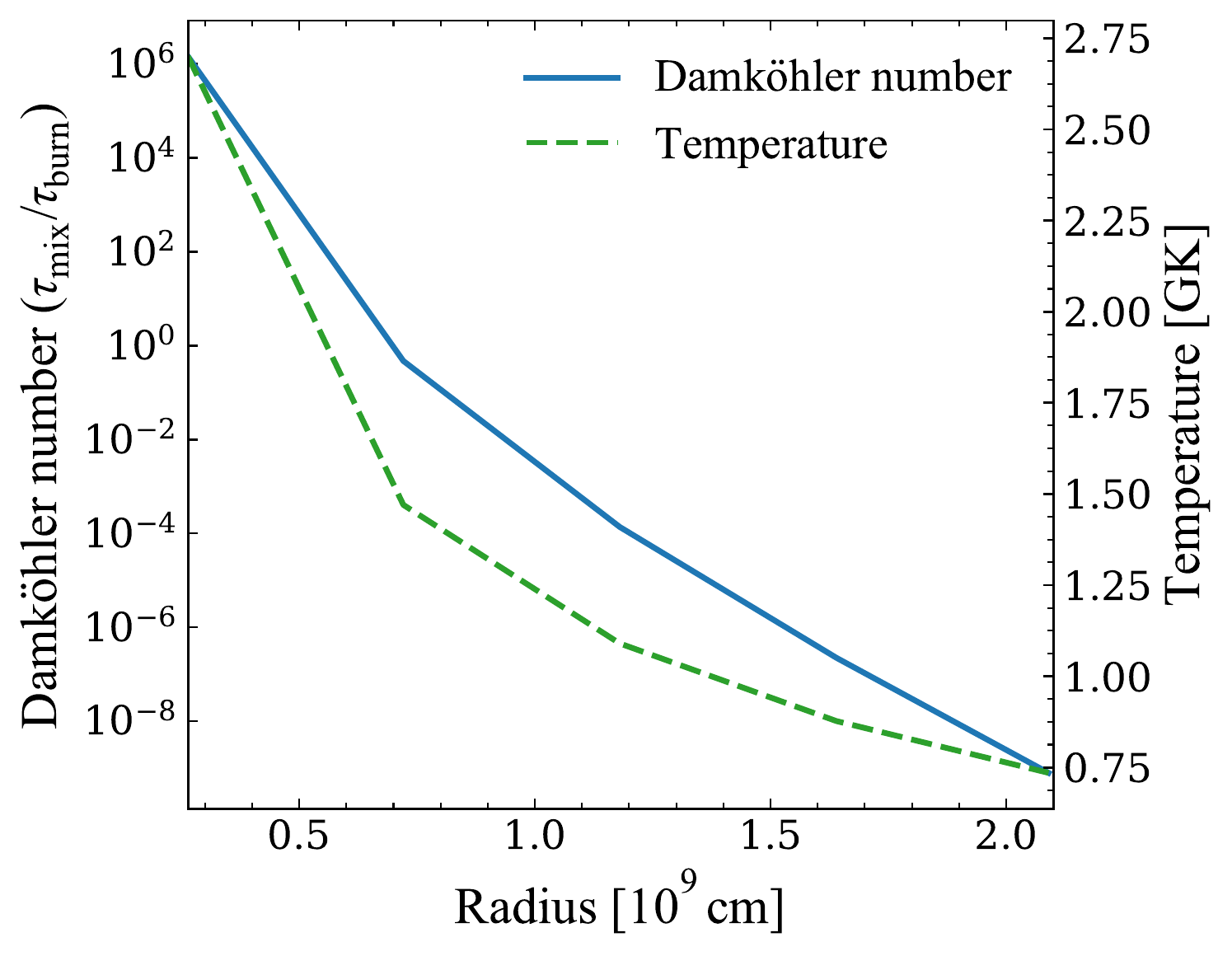}
	\caption{Damk\"ohler number (blue solid line) and temperature (green dashed line) as a function of radius within the Si shell, about 20 cycles before the Si-C shell merger. The range of radius shown represents the boundaries of the convective Si shell.
}
\label{fig:da_vs_r}
\end{figure}

We have estimated the burning time scale of $^{12}$C by performing a simple
nuclear network calculation beginning from 90\,\% $^{28}$Si and 10\,\% $^{12}$C. We
keep the temperature fixed at 3~GK and the density at
$1.4\times10^8$~g~cm$^{-3}$. We define the burning time scale as the e-folding
time of the $^{12}$C mass fraction $X_\mathrm{C}$ under such conditions,
\begin{equation}
	\tau_\mathrm{burn}\approx\frac{\mathrm{d}t}{\mathrm{d~ln}X_\mathrm{C}}
	\quad .
\end{equation}
For example, near the bottom of the Si shell  the time scale is $\sim\,10^{-3}$\,s, which is much shorter than the
$\sim\,10^3$\,s convective time scale, giving an exceptionally large Damk\"ohler
number of Da~$=\tau_\mathrm{mix}/\tau_\mathrm{burn} \approx 10^6$ (Figure~\ref{fig:da_vs_r}).
This means the material in the C shell will never
actually reach the bottom of the convection zone. Instead, the situation is reminiscent of the H-ingestion into He-shell convection in a post-AGB star. The distributed combustion flame is located where Da~$\sim\,1$ \citep{Herwig:2011dj}, which in this case is in the lower third of the Si convection zone. Depending on the energy release of the convective-reactive burn relative to the initial convective kinetic energy, a global non-spherical convective-reactive instability may ensue, such as the Global Oscillation of Shell H-ingestion (GOSH) that has been reported for H-ingestion into He-shell flash convection \citep{Herwig2014a}. The exact nature of the convective-reactive event and its impact on yields requires a more comprehensive 3D hydrodynamics and nucleosynthesis simulation approach which is beyond the scope of this work.  

\section{Conclusions}
\label{sec:conclusions}

In this paper, we analyzed the Si-C shell merger occurring in the 20\,$M_\odot$ model at $Z=0.01$ of the NuGrid collaboration (\citealt{2018MNRAS.480..538R}).
The Si-C shell merger occurs roughly an hour before collapse. Following this
event, a large amount of incomplete Si-burning material, including $^{51}$V and $^{52}$Cr, is
mixed all the way from the Si core to a mass coordinate of 5\,\msun, which
represents the upper boundary of the C shell as it was before the merger event.
The convective time scale of this mixed shell is similar to the delay before the
star collapses. As a first order approximation, it is therefore possible for the
incomplete Si-burning material to mix and fill the Si-C shell by the time of the explosion. Because the adopted mass cut for this model is 2.77\,\msun, a significant fraction of Cr is ejected during the explosion.

Using our GCE code \texttt{OMEGA+} (\citealt{2018ApJ...859...67C}), and assuming
that the ejecta of this specific 20\,$M_\odot$ model at $Z=0.01$ is
representative of all 20\,$M_\odot$ stars at that metallicity, we overestimate
the predicted evolution of [Cr/Fe] in the Milky Way by almost an order of
magnitude at [Fe/H]\,$\sim$\,0. A question that emerges is whether or not Si-C shell mergers occur in nature. From this experiment, the only conclusion we can draw from a GCE perspective is that the specific conditions (assumptions), in which this 20\,$M_\odot$ model evolves and explodes, cannot be representative of all 20\,$M_\odot$ stars and should be extremely rare if they occur at all.

From the analysis of the stellar evolution model, the Si-C merger could launch a non-spherical, global, convective-reactive instability similar to the GOSH found in H-ingestion in post-AGB stars \citep{Herwig2014a}. Such an instability could seed substantial non-spherical perturbations of the initial conditions for the SN explosion. Another implication could be that such an instability enhances mixing of Cr-rich Si-shell material into the C-shell above. This would impact the amount of Cr ejected in this model.
If such instabilities occur, their properties will depend on the detailed balance between energy produced from the entrainment of C-shell material and the driving energetics of the Si-burning convection shell. Without such instability, if no or only a partial merger would take place, the convective mixing time scale in the C-shell is similar to the remaining time to collapse, and dredge-up of Cr into the C-shell convection zone would likely be incomplete. 

3D hydrodynamic simulations are required to investigate the range of mixing and burning conditions during interactions between the Si and C shells. Such an investigation would address important questions. How would a Si-C shell merger look in multi-dimensional hydrodynamic simulations? To what extent would Cr make its way up into the C shell? How would an interaction between the Si and C shells impact the pre-supernova structure and the dynamics of the supernova explosion? Would the star explode? In any case, GCE would provide stringent constraints on the frequency and efficiency of Cr production through this process, which this study shows must remain small.

\acknowledgments
This research is supported by the ERC Consolidator Grant (Hungary) funding scheme (Project RADIOSTAR, G.A. n. 724560), by the National Science Foundation (NSF, USA) under grant No. PHY-1430152 (JINA Center for the Evolution of the Elements), and by the US Department of Energy LDRD program through the Los Alamos National Laboratory. Los Alamos National Laboratory is operated by Triad National Security, LLC, for the National Nuclear Security Administration of U.S. Department of Energy (Contract No. 89233218NCA000001). MP acknowledges significant support to NuGrid from STFC (through the University of Hull's Consolidated Grant ST/R000840/1), and access to {\sc viper}, the University of Hull High Performance Computing Facility. MP acknowledges the support from the "Lendulet-2014" Programme of the Hungarian Academy of Sciences (Hungary).

\vspace{5mm}

\bibliographystyle{yahapj}
\bibliography{main_arxiv}

\end{document}